\documentclass[preprint,12pt]{elsarticle}

\usepackage{courier}
\usepackage[ngerman, russian, english]{babel}
\usepackage[T2A]{fontenc}
\usepackage[utf8]{inputenc}  
\usepackage{hyperref, url}

\usepackage{hypdvips}

\usepackage{epsfig, subfigure}
\usepackage{float}
\usepackage{pst-plot, pstricks}
\usepackage{wrapfig}

\usepackage{fancyhdr}
\usepackage{ae}
\usepackage{tikz}

\usepackage{amsmath, amssymb, amsfonts, wasysym, mathrsfs}
\usepackage{latexsym, eurosym}

\usepackage[safe]{tipa}
\usepackage{multirow, multicol, sgame}
\usepackage{pbsi, textcomp}

\newcommand{\Tab}[1]{Tab.\ref{#1}}

\newcommand{\kA}{\selectlanguage{russian}Ы\selectlanguage{english}}
\newcommand{\kB}{\selectlanguage{russian}Э\selectlanguage{english}}
\newcommand{\kC}{\selectlanguage{russian}Ю\selectlanguage{english}}
\newcommand{\kD}{\selectlanguage{russian}Я\selectlanguage{english}}

\journal{Artificial Intelligence}

\begin{document}

\selectlanguage{english}

\begin{frontmatter}



\title{If more than Analytical Modeling is Needed to Predict Real Agents' Strategic Interaction}


\author[rt]{Rustam Tagiew}

\ead{tagiew@informatik.tu-freiberg.de}

\address[rt]{Institute for Computer Science, TU Bergakademie Freiberg, Bernhard-von-Cotta-Str. 2, 09596 Freiberg, Germany\\
Phone: +49 (0)3731 / 39 3113\\
Fax: +49 (0)3731 / 39 2298}

\begin{abstract}
This paper presents the research on the interdisciplinary research infrastructure for understanding human reasoning in game-theoretic terms. Strategic reasoning is considered to impact human decision making in social, economical and competitive interactions. The provided introduction explains and connects concepts from AI, game theory and psychology. First result is a concept of interdisciplinary game description language as a part of the focused interdisciplinary research infrastructure. The need of this domain-specific language is motivated and is aimed to accelerate the current developments. As second result, the paper provides a summary of ongoing research and its significance.
\end{abstract}


\end{frontmatter}

\section{Introduction}
\label{k.intr}
\indent There is more than one scientific discipline that can be used as a source for predicting outcomes of human social, economical and competitive interactions on the granularity level of individual decisions \cite[p.4]{tagiewPhD}. For autonomous intelligent systems, which perceive, decide and act in an environment according to their preferences, the notion {\it agent} can be used \cite[chap.2]{russel}. In addition to computer science related disciplines (AI and MAS), this notion is also used in sociology regarding people \cite{soz}. Further, people and artificial agents are called as {\it real agents} in contrary to theoretical concepts.\\
\indent Game theory is the appropriate mathematical discipline, if the agents' interactions can be considered as {\it strategic} \cite{rubin}. In strategic interactions, agents are {\it rational} and apply mutually and even recursively the concept {\it rationality}. This concept justifies not only decisions, but also predictions of outcomes. A rational agent makes always decisions, whose execution has according to his subjective estimation the most preferred consequences for him \cite{russel,rubin}. The informative value and the correctness of the subjective estimation depend on the level of agent's intelligence.\\
\indent A rational agent is not required originally to be familiar with the concept ``rational agent'' appellative for him. A rational agent is a {\it strategic agent} \cite[p.109]{tagiewPhD}, if the case of familiarity with the nested and recursive application of the concept ``rational agent'' is given. For strategic agents Alice and Bob e.g., nested rationality means that Alice predicts Bob's prediction of her rational decisions, if she knows his present preferences over the consequences of her actions. Therefore, the strategic agents are a subset of the rational agents. Strategic agents with unbounded intelligence satisfy the definition of {\it game-theoretic agents} \cite{gametheoretic}.\\
\indent If interacting strategic agents identify each other as strategic agents, it is called a {\it strategic interaction}. Further, {\it game} is a notion introduced by von Neumann and Morgenstern \cite{neumann} for the formal structure of a concrete strategic interaction. Although the modern game-theoretic textbooks contain mostly fictive scenarios, the reader is asked to resist the conclusion that real-life cases can not be formalized as games. In contrast, early research on game theory was forced by the RAND corporation in order to provide predictions for the outcome of Cold War \cite{randgt}. The fictiveness of game-theoretic examples facilitates rather an objective and politically inoffensive scientific discussion. And as conjectured, formalized rules of parlor games like chess are also games in game-theoretic terms.\\ 
\indent Agents involved in a game are called {\it players}. A game in {\it normal form} consists of the {\it strategies} (sets of decisions) and players' preferences over outcomes \cite{rubin}. {\it Finite normal form games} contain a finite number of outcomes. Solving a game in game-theoretic terms means to provide the {\it equilibria}. Due to its definition, an equilibrium is an irrevocable combination of players' strategies -- none of the players can improve his outcome by altering his chosen strategy. If players' preferences are defined by their {\it payoff functions}, the equilibria of finite normal form games are guaranteed and called Nash equilibria \cite{nash}. A payoff or also utility function assigns a numeric value as magnitude of preference to every outcome. Calculation of equilibria is generally NP-hard \cite{npcompnash}. {\it GAMBIT} is a library of implemented algorithms for this task \cite{gambit}.\\  
\indent In the case of players, who are intelligent enough to solve their game, an equilibrium should (immediately) occur independently of players' builds or even real identities. This is fundamentally different to the common AI approach, where programs compete the last decades in playing chess e.g. still without achieving any irrevocable solution \cite[chap.5]{russel}. The common AI approach is reasonable nevertheless. Although the existence of an equilibrium is proven for chess, analytical abilities of none of the present game theorists suffices to show at least, whether the equilibrium implicates the win of the whites \cite{stratsp}. Therefore, an equilibrium is not guaranteed to occur, if at least one of the players is not able to solve the game. The AI approach is referenced further as {\it game playing}.\\
\indent Formalization of a concrete strategic interaction as a game is not trivial, if explicit rules lack contrary to such cases like chess. This step is mostly done manually by the game theorists themselves, before a game can be solved. (Manual) game-theoretic formalization is not guaranteed to be accurate and mostly results in far too simplified {\it ``toy games''} as criticized in AI literature \cite{sequenceform}. A strategic agent faces the same problem as a strict outsider like a game theorist. He has to answer the question correctly, which interacting (strategic) agents really exist in the environment and in what kind of game he is involved in. It is called {\it incomplete information} in game-theoretic terms \cite{rubin}, if this question can not be clarified.\\
\indent In the case that a probability distribution over all possible game-theoretic formalizations can be provided, incomplete information can be transformed to {\it imperfect information} as proposed by Harsanyi \cite{jharsanyi}. For instance, poker is a game of imperfect information, where every player is supposed to be aware about the probability distribution over the possible hands, but does not know the current hands. It is not obligatory that a game becomes a {\it common knowledge} among its players -- it is also possible that a player bears his own variant of the game in mind, he misconceives to be involved in. As a reminder, common knowledge is that, what everybody knows and everybody knows that everybody knows it and prepending ``everybody knows that'' ad infinitum \cite{FHMV}.\\
\indent Common knowledge of the game would not exist during a poker round, if Alice hides cards in her sleeve unknown to her opponent Bob. Equilibria can be calculated for the original game as well as for the ``cheated'' game. Alice considers that Bob's actions should conform the equilibria calculated in the original game, but does not act accordingly to the equilibria of the ``cheated'' game -- she rather calculates the equilibria of a {\it global game} consisting of the original game and the ``cheated'' game \cite{globalgame}. Generally, players may have an intractable number of mutual nested believes about the details of their game. For instance, Alice could reckon to some extent that Bob also cheats, Bob could suspect Alice of considering him as a cheater and so on. An intractable number of mutual nested believes results in an intractable size of the global game, where no game-theoretic solution is guaranteed to be calculated or even to exist in the case of infinity. Further, there is no trivial general way for direct control or prediction of the players' mutual nested believes and their development. Therefore, cases are concentrated on, where a game of imperfect information is its players' common knowledge.\\
\indent This paper concentrates on the relevance of strategic reasoning in real agents' decision making, which causes an (immediately) occurring equilibrium, if the unbounded intelligence is assumed. In the cases of absent strategic reasoning, game theory can still predict the direction of convergence id est towards an equilibrium. This is proposed by Price \cite{MaynardSmith1973} for prediction of stochastic processes, more precisely populations in biological terms, and is called {\it evolutionary game theory} \cite{biogame}. A convergence is also observed in the case of using {\it reinforcement learning} instead of strategic reasoning \cite[e.g.]{backgammon}. The assumption of this paper is that the usage of the concept strategic reasoning can be extended by newer methods. This assumption does not negate the existence of reinforcement learning e.g. in real agents' strategic interactions and is discussed next in the case of people.\\
\indent Although the real human preferences are a subject of philosophical discussions \cite{humannat}, the application of strategic reasoning assumes that they can be captured in concrete interactions as required for modeling rationality. The consideration of people as rational agents is disputed at least in psychology \cite[pp.527--530]{hrational}, where even a scientifically accessible argumentation exposes the existence of stable and consistent human preferences as a myth \cite{sociokritik}. Since the last six decades nevertheless, the common scientific standards for {\it econometric} experiments are that subjects' preferences over outcomes can be insured by paying differing amounts of money \cite{performmoney, experimental}. However, insuring preferences by money is criticized by the term {\it homo economicus} as well \cite{evolving}.\\
\indent The ability of identifying other agents and of modeling their reasoning corresponds with the psychological term {\it ToM} (Theory of Mind) \cite{verbrugge}, which lacks almost only in the cases of autism. For application of strategic reasoning, subjects as well as researchers, who both are supposed to be non-autistic people, may be then able of modeling of others' strategic reasoning too. In Wason task at least, subjects' reasoning does not match the researchers' one though \cite{wason}. People may use no logic at all \cite{nonlogic}, but also mistake seriously in the calculus of probabilities \cite{tversky}.\\
\indent The data of econometric experiments does not match the equilibria of games according to which they are conducted \cite{poolgame,vspt}. That means that the strategic reasoning according to the global (researchers' point of view) game does not arise among the subjects due to a set of reasons, which should be clarified. There is a need for more than only the pure analytical game theory, because even people familiar with game theory are observed to deviate from equilibria in multiple cases \cite{vspt}. It is gathering and analyzing of data from experiments and field studies. This paper steps further -- it proves the potential of making the interdisciplinary research on real agents' strategic interactions more efficient. Like in bio-informatics \cite{forsinfr}, it is supposed to be done by an {\it interdisciplinary research infrastructure} -- domain specific languages and common tools.\\
\indent The paper is organized as follow. Next section summarizes the main concepts for the interdisciplinary research infrastructure. Then, the section \ref{k.ongoing} presents detailed the hitherto research. At the end, the results are concluded in order to figure out the remaining construction sites.\\
\section{Conceptualizing Interdisciplinary Research Infrastructure}
\label{k.rinf}
\indent A conceptualization of the already partially existing (interdisciplinary) research infrastructure for real agents' strategic interactions follows. It aims to provide an elaborate overview and an exhaustive motivation. Artificial agents are also included into consideration, although this paper concentrates mostly on people. People can be replaced by artificial agents in order to simulate or to intervene human strategic interactions. Whether simulation or intervention -- in both cases, artificial agents can cut costs, allow a direct control of their builds, can be numerously deployed and are almost unlimited in period of use.\\
\indent In order to avoid incomplete information, which can make the application of strategic reasoning intractable, an explicit formalization is to be used. Because a formalization of a concrete strategic interaction can be inaccurate, it is reasonable to execute it inversely. A concrete strategic interaction has to be created out of an already existing game. This is {\it game realization} and a software-based game realization is a {\it game implementation} \cite[p.108]{tagiewPhD}. Game realization is the almost always unmentioned action after {\it mechanism design} \cite[p.632]{russel}. Mechanism design is inverse to game solving -- adjusting of a game to already predetermined desired equilibria. If game realization is impossible, mechanism design is futile.\\
\indent In real-life cases, games can be realized by physical conditions, by non-participating agents, by participating agents themselves or by a subset of the $3$ previous instances \cite{tagiewggma}. If participating agents are responsible for a (partial) game realization, they should prefer the compliance with rules over the advantages from ``cheating''. A non-participating agent responsible for game realization can be modeled as rational too. For instance, the attorney in {\it prisoner's dilemma} keeps his word in order to not risk his reputation, where prisoner's dilemma is supposed to be familiar to the reader.\\
\indent In the case of game implementation, the software can be divided into fractions: {\it game management}, game-solving algorithms, game-playing algorithms and auxiliary algorithms. Game management is the part of software, which executes the rules and calculates outcomes \cite{ggp}. It may also record in order to gather data \cite{framasi01}. Human-computer interfaces aka {\it proxy agents} \cite{aviNegot} are examples of auxiliary algorithms. Additionally, game implementation minus game-solving/playing algorithms is called {\it game infrastructure} \cite[p.53]{tagiewPhD}.\\
\indent In the case of explicit rules, the question about the form of games arises. The normal form is one of the two most general forms to express games in {\it non-co-operative game theory}. In {\it co-operative game theory} or also {\it coalitional game theory} \cite{coopgt}, players may covenant and group into {\it coalitions}. The process of negotiating and the way of ensuring the agreements themselves are not issues of co-operative game theory. From non-co-operative game-theoretic point of view, disadvantages should arise for the players, who break the agreements. From both points of view, a player makes a rational decision, whether it is his own behavior or an agreement about a co-operative behavior. Both points of view are considered to be equal \cite{rubin}. Due to the fact that the pure analytical game theory does not suffice, the interest is focused a more detailed form -- the {\it extensive form} -- the second most general to express games in non-co-operative game theory. In contrast to others, the extensive form captures separately the actions' sequences and their alternating subsequences in a representation known as {\it game tree} in AI. Therefore, the actions' sequences can be called as {\it root paths}. The normal form and the coalitional formalization are skipped as argued because of their higher abstraction. Also skipped is the consideration of continuous games \cite{differen}, where equilibra can be acquired by solving of differential equations and no general form exit.\\
\indent The theoretical appropriateness of the extensive form does not result a computational one -- the problem is the inappropriate size of a games represented in extensive form. For instance, one can consider that the rules of chess are sent as a game tree via network. It is rather an {\it Interdisciplinary Game Description Language} (IGDL) having expressive power of the extensive form, which is needed for $4$ rough categories of considered instances within and beyond game implementations. These categories are 
\begin{center}\begin{minipage}[]{10cm}\begin{description}
 \item[\kA)] game-solving algorithms,
 \item[\kB)] game-playing algorithms,
 \item[\kC)] non-solving/playing algorithms (game management, mechanism design, auxiliaries etc.) and
 \item[\kD)] scientific human users (sociologists e.g.).
\end{description}\end{minipage}\end{center} Less important issues are skipped out of this consideration for IGDL -- issues like adjustments for {\it evolutionary mechanism design} \cite{coevolution} and for non-scientific human users. Reducing the size compared with an equivalent representation in extensive form is called further {\it compactness} \cite[p.65]{tagiewPhD}. Compactness is not the only criterion for IGDL. For the categories \kA--\kD, one can summarize the partial criteria as following:
\begin{enumerate}
\item \textbf{Computational speed-up.} Regularities like symmetries can be used in order to reduce the computation time of equilibria \cite{succinct}. This feature is captured by formalisms called {\it succinct games} and should be also provided by IGDL. Reduction of computation time by using compact representation applies also to game-playing algorithms \cite[e.g.]{flux} and can be considered in the case of game management.
\item \textbf{Re-usability \& comparability.} A language for games forces the re-usability and also comparability of game-playing algorithms as suggested by Pell \cite{pellfirst}. This applies also for game-solving algorithms \cite{gambit} and for non-solving/playing algorithms algorithms \cite{zTree}.
\item \textbf{Interdisciplinary human usability.} IGDL should prevent the scientific manual game-theoretic formalization from resulting in ``toy games'' as criticized in AI literature \cite{sequenceform}. A graphical representation of a game may improve the usability even more \cite{tagiewcimca}. 
\item \textbf{Decidability.} This feature should be provided for IGDL in order to ensure that the calculation of outcomes definitely terminates \cite{ggp08}. This also important for game-solving/playing algorithms to be able to calculate the consequences of actions.
\item \textbf{General compact interchange format.} For interfacing instances of the categories \kA--\kD, IGDL should satisfy the need of a compact interchange format. At same time, IGDL should be as general as possible, where the facility to express n-person games of imperfect information is most general. Finally, instances of all the mentioned categories should be at least theoretically IGDL-compatible in order to skip reformatting, id est to facilitate their efficient mutual integration. 
\item \textbf{Time.} Time remains the issue disregarded by the extensive form. As one can conceive by comparing game playing in fast chess and in normal chess, time given for making decisions impacts them. Therefore, time is needed to be included in IGDL in order to ascertain the time dependent details by the explicit rules. Otherwise, the durations of actions' sequences e.g. may depend on the current game implementation and not be given explicitly in advance conjoined with the game. 
\end{enumerate}
\indent Some examples for usage of IGDL can be provided. As 1st example, IGDL-compatible chess playing algorithms can be incorporated into system, which compete in playing other chess-like games described in IGDL. As 2nd example, a IGDL-based game editor can be used to allow non-computer scientists to set-up their own experiments. As 3rd example, data gathered in experiments conducted according to a game described in IGDL can be compared with the equilibria calculated by IGDL-based game-solving algorithms for the same game. As 4th example, game described in IGDL can be easily forwarded to the IGDL-based game-playing algorithms for an approximative solution, if IGDL-based game-solving algorithms fail to output timely. As 5th example (proposed by \cite{gaechter}), the data of the experiments conducted based on IGDL can be better compared or even stored in a central web database like the state of art in bio-informatics. Games described in a special language with and without conjoined time dependent details are called further {\it game descriptions}.\\
\section{Summarizing Ongoing Research}
\label{k.ongoing}
\begin{table}
\begin{center}
\caption{Overview of the precursors for IGDL \cite[p.69]{tagiewPhD}. Numbers in the \textbf{Crit.}-subrow are the satisfied criteria from the section \ref{k.rinf}, The \textbf{Used}-subrow shows the categories of used software. The \textbf{Means}-row includes the rough categories of the means used for describing games. ``perfect'' concerning information is to be inferred in case of missing ``imperfect'' and ``simultaneous moves''.}\label{t.gamelanguages}\begin{tabular}{l|c|l|l}
\hline
\textbf{Approach}             & \textbf{Crit.}     & \textbf{Means} & \textbf{Class of games} \\
\textbf{Citation}             & \textbf{Used}      &                & \\
\hline
congestion games              & 1,4                & functions      & subset of n-person games,  \\
\cite{congest}                & \kA,\kC            &                & simultaneous moves\\
\hline
sequential form               & 1,4                & matrices       & 2-person games of  \\
\cite{sequenceform}           & \kA,\kC            &                & imperfect information \\
\hline
graph games                   & 1,4                & graphs,        & n-person games,  \\
\cite{grgames}                & \kA,\kC            & functions      & simultaneous moves \\
\hline
local effect games            & 1,4                & functions      & subset of n-person games,  \\
\cite{localeff}               & \kA,\kC            &                & simultaneous moves\\
\hline
action graph games            & 1,4                & graphs,        & n-person games,  \\
\cite{agg}                    & \kA,\kC            & functions      & simultaneous moves      \\
\hline
Gala                          & 2                  & logic          & n-person games of  \\
\cite{gala}                   & \kA                &                & imperfect information   \\
\hline
MAID                          & 1,4                & Bayes-nets     & n-person games of  \\
\cite{maidfirst}              & \kA,\kB            &                & imperfect information  \\
\hline
continuous games              & 6                  & functions      & subset of 2-person games of  \\
\cite{contin2007}             & \kA                &                & imperfect information\\
\hline
timed games                   & 6                  & functions      & 2-person games \\
\cite{timedautomaton,gamecontrolltheory1} & \kB    &                & \\
\hline
GDL                           & 1,2,4              & logic          & deterministic n-person games,\\
\cite{ggp08}                  & \kB,\kC            &                & simultaneous moves  \\  
\hline
GDL-II                        & 1,2,4,5            & logic          & n-person games of  \\
\cite{gdl2}                   & \kB,\kC            &                & imperfect information  \\ 
\hline
game Petri-nets               & 1                  & Petri-nets     & deterministic n-person games,\\
\cite{gamepetri}              & \kA                &                & simultaneous moves \\
\hline
PNSI                          & 2,4--6             & Petri-nets     & n-person games of  \\
\cite{tagiewiccci}            & \kA--\kC           &                & imperfect information \\
\hline
SIDL2.0                       & 2,5,6                & logic          & n-person games of  \\
\cite[p.98]{tagiewPhD}        & \kC                &                & imperfect information  \\
\hline
z-Tree-language               & 2,3,5,6            & imperative     & n-person games of  \\
\cite{zTree}                  & \kC,\kD            & language       & imperfect information  \\
\hline
\end{tabular}
\end{center}
\end{table}
\indent IGDL is the desired domain-specific language, which has already some precursors and these precursors are summarized in this section. It is also possible that the ongoing research will bear different concurrent versions of IGDL. In order to assess the hitherto approaches better, a rough categorization of the used formal means is provided. Such categories are functions, graphs, logic, Petri-nets and so on. \Tab{t.gamelanguages} contains the regarded approaches and their rough categories. The ability to describe {\it simultaneous moves} (id est actions) is subset to more general imperfect information, because other players' actions can be unobserved only during a simultaneous execution. In {\it deterministic} games, it is impossible to describe a probability distribution over possible subsequences of actions.\\
\indent In discrete non-co-operative game theory, there are different approaches for compact game forms, whose aim is computational speed-up \cite{succinct}. The most important of them chronologically ordered are {\it congestion games} \cite{congest}, {\it sequential form} \cite{sequenceform}, {\it graph games} \cite{grgames}, {\it local effect games} \cite{localeff} and {\it action graph games} \cite{agg}. The software {\it GAMUT} can generate random games of these and other kinds, where the extensive form is included \cite{gamut}. Computational speed-up of sequential form in solving 2-person-games of imperfect information is used in GAMBIT \cite{gambit}.\\
\indent {\it GAme LAnguage} (Gala) \cite{gala} is developed in order to provide an interface to the game-solving algorithms. Factually, Gala affords Prolog-based game descriptions, where a game of extensive form or of normal form can be generated. The generated game-theoretic representation can be forwarded to GAMBIT or other game-solving algorithms. The main improvement of Gala compared to these representations is the game descriptions' compactness as perceived at least at the examples from Gala's software package. Due to full-scale Prolog, Gala does not provide decidability. Therefore, the generation of extensive form games from Gala game descriptions must not terminate.\\
\indent There is a subset of game-playing algorithms, which is not only aimed to play games, but also to simulate human (strategic) reasoning in them. Simulating human reasoning falls in the subject of {\it cognitive science}. There are currently two different approaches, where the human strategic reasoning has to be expressed in a general language in order to facilitate an efficient comparability of the models. The first is based on {\it cognitive architectures} \cite{wallach, westlebierelast}, which are languages for models of general human reasoning. The second is based on {\it Multi-Agent Influence Diagrams} (MAID) \cite{maidfirst}. The second factually conforms the game-theoretic point of view on strategic interactions and provides the alternative language MAID for describing games. MAID are shown to be expressive enough to represent n-person games of imperfect information in the algorithms for game playing. MAID can be also transformed to extensive form in order to solve them \cite{maidfull}.\\
\indent The previously discussed literature does not mention the inclusion of time dependent details in game descriptions, what some theoretical approaches from game theory, {\it concurrency theory} \cite{gamecontrolltheory2} and {\it control engineering} \cite{gamecontrolltheory1} aim. A current work  \cite{contin2007} in game theory extends extensive form games of perfect information and continuous time \cite{contin1989} to such of imperfect information -- {\it continuous games}. In \cite{contin1989, contin2007}, only a subset of such games is regarded -- the set of player's actions is always the same. Generally, a point of time is assigned to every action and time grows strictly over a sequence of actions.\\ 
\indent {\it Game Description Language} (GDL) successed in sparking an international programing competition on general game playing \cite{ggp08, gdlweb}. A concrete game description in GDL is sent to an artificial game player and never to its human programmer -- the programmer knows only the structure of GDL. This satisfies the criterion 2. For the criterion 5, generality of IGDL's precursors is a trend -- either it will be possible to describe n-person games of imperfect information in IGDL or IGDL should be extended to facilitate that. This trend caused the development of GDL-II \cite{gdl2}, which is an extension of GDL for n-person games of imperfect information. Like GDL, GDL-II is based on Datalog, which is a version of Prolog \cite{datalogcom}. Datalog guarantees decidability by banning functions, limiting variables' ranges and restricting recursion. Due to the decidability, the existing game management algorithm based on GDL-II is guaranteed to terminate. However, the ban of functions worsen compactness. For instance, if arithmetic addition is needed to describe actions' consequences in a game, the result for every required summands' combination must be separately defined in the game description.\\
\indent There are no time dependent details included explicitly in GDL-II game descriptions. GDL and GDL-II describe games in a way {\it STRIPS-like} \cite{strips} languages do. In languages for planning tasks, STRIPS-like descriptions can be replaced by descriptions based on Petri nets \cite{petristrips}. {\it Petri Nets for Strategic Interaction} (PNSI) is a game description language, which is proposed chronologically between GDL and GDL-II \cite{tagiewcimca}. PNSI uses basic Petri nets instead of logic. Petri nets are also known being used in game theory to describe subclasses of games \cite{gamepetri}. PNSI provides decidability \cite[p.89]{tagiewPhD}. The advantages of PNSI compared with GDL-II are the graphical representation of Petri nets and the ability to describe games of {\it equidistant time}. Equidistant time means that the game management algorithm for PNSI pauses exactly for one {\it chronon} between two {\it game states} \cite{tagiewiccci}, where a game state is also a node of its game tree. In this context, a chronon is a constant period of time, which is explicitly known to players. During a chronon, players' actions can be submitted. The game management algorithm for GDL-II is not of equidistant time, because the next state is calculated exactly after the submission of the last action, if it is inside the allowed time period. The time point of the last submission may vary depending on players. Of cause, GDL-II is supposed be also able to describe games of equidistant time, if its game management algorithm is modified as proposed for PNSI.\\
\indent For PNSI, there exist an algorithm that can generate games of extensive form from game descriptions as in the case of Gala \cite{tagiewiccci}. Therefore, PNSI provides an interface to game solving algorithms. A game of extensive form generated based on a PNSI game description is a slightly modified {\it state transition system} of the underlying basic Petri net. In this context, a state transition system is an oriented graph consisting of game states, where every edge represents a  progression in time. There is still no algorithm for GDL-II to generate games of extensive form. A state transition system can be also generated for GDL-II game descriptions \cite{gdl2}. Therefore, generation of extensive form games is supposed to be also possible for GDL-II game description.\\
\indent PNSI suffers of insufficient compactness as well as GDL-II but in a different way. The arithmetic addition and subtraction are banned in Datalog, but inherently supported by basic Petri nets. On the other hand, basic Petri nets require every state to be decoded as a vector of natural numbers and do not have other operations than the addition and the subtraction. The game description of the parlor game {\it Nim} needs in PNSI much less space than in GDL-II \cite{tagiewPhD}. The opposite for chess, a GDL-II chess description needs less.\\
\indent Dropping the criterion of decidability may dramatically improve compactness. {\it Strategic Interaction Definition Language} (SIDL) is based on ISO-Prolog, does not provide decidability and attains for example games a higher compactness \cite[p.98]{tagiewPhD}. Of cause, the widely used game description in an imperative language can be also mentioned. However, the game management part of software aka {\it game server} is then required to send its own code in order to provide explicit rules \cite[p.54]{tagiewPhD}.\\
\indent For scientific human users beyond computer science, there is an ongoing development of user-friendly software for experiments \cite{teec}. {\it RatImage} \cite{ratimage} and {\it TEEC} \cite{teec} are examples of the first generation of such software. They are libraries, which facilitate programming. The second generation provides already domain-specific languages for the game management and the layout of human-computer interfaces. These are {\it ComLabGames} \cite{comlabgames} and {\it z-Tree} \cite{zTree}. z-Tree is the most used \cite{gaechter}. There is a z-Tree-language, which is actually an imperative language, in which the game and also the human-computer interfaces can be described. This language does not provide decidability. It has no relations to game solving or playing algorithms.\\
\section{Conclusion}
\indent A large-scale view on the problem of understanding human strategic reasoning is presented. The elaborated solution is the development of the interdisciplinary research infrastructure. This research infrastructure is proposed to make the interdisciplinary research more efficient, as it is already observed in similar interdisciplinary problems. As an underline of the large-scale view, there are matters chained from different sources, which have been never cited together before. For instance, GDL-II and z-Tree are such matters.\\
\indent The elaborated concept is the domain-specific language IGDL. A summary of its precursors shows that none of these is developed enough to satisfy IGDL's full outline. There is still no language, which incorporates the graphical representation like PNSI, compactness improvements like GDL-II and a proven human usability like z-Tree-language.\\
\section{References}
\bibliographystyle{elsarticle-num}
\bibliography{tagiew}
\end{document}